# SOME NEW MODELS FOR STRANGE QUARK STARS WITH ISOTROPIC PRESSURE

**Manuel Malaver***

*Universidad Marítima del Caribe, Departamento de Ciencias Básicas, Catia la Mar, Venezuela
(Email: mmf.umc@gmail.com)

**Abstract**: We found new class of solutions to the Einstein-Maxwell system of equations for charged quark matter within the framework of MIT Bag Model considering a gravitational potential Z(x) proposed by Thirukkanesh and Ragel (2013), which depends on an adjustable parameter n. Variables as the energy density, charge density, pressure and the metric functions are written in terms of elementary and polinominal functions. We show that the form chosen for the gravitational potential allows obtain physically acceptable solutions with any value of the adjustable parameter.

## 1. Introduction

From the development of Einstein´s theory of general relativity, the description of compact objects has been a central issue in relativistic astrophysics in the last few decades [1,2]. Recent experimental observations in binary pulsars [2] suggest that could be quark stars. The existence of quark stars in hydrostatic equilibrium was first by Itoh [3] in a seminal treatment. Recently, the study of strange stars consisting of quark matter has stimulated much interest since could represent the most energetically favourable state of baryon matter.

The physics of ultrahigh densities is not well understood and many of the strange stars studies have been performed within the framework of the MIT bag model [4]. In this model, the strange matter equation of state has a simple linear form given by $p = \frac{1}{3}(\rho - 4B)$ where $\rho$ is the energy density, p is the isotropic pressure and B is the bag constant. However, in theoretical works of realistic stellar models [5-8] it has been suggested that superdense matter may be anisotropic, at least in some density ranges.

Many researchers have used a great variety of mathematical techniques to try to obtain exact solutions for quark stars within the framework of MIT bag model, since it has been demonstrated by Komathiraj and Maharaj [9], Malaver [10], Thirukkanesh and Maharaj [11] and Thirukkanesh and Ragel [12]. Feroze and Siddiqui [13] and Malaver [14] consider a quadratic equation of state for the matter distribution and specify particular forms for the gravitational potential and electric field intensity. Mafa Takisa and Maharaj [15] obtained new exact solutions to the Einstein-Maxwell system of equations with a polytropic equation of state. Thirukkanesh and Ragel [16] have obtained particular models of anisotropic fluids with polytropic equation of state which are consistent with the reported experimental observations. More recently, Malaver [17,18] generated new exact solutions to the Einstein-Maxwell system considering Van der Waals modified equation of state with and without polytropical exponent. Mak and Harko [19] found a relativistic model of strange quark star with the suppositions of spherical symmetry and conformal Killing vector.

Our objective in this paper is to generate a new class for charged isotropic matter with the bag equation of state that presents a linear relation between the energy density and the radial pressure in static spherically symmetric spacetime using a gravitational potential Z(x) of Thirukkanesh and Ragel [12] which depends on an adjustable parameter n. We have obtained some new classes of static spherically symmetrical models of charged matter where the variation of the parameter n modifies the radial pressure, charge density, energy density and the metric functions of the compact objects. This article is organized as follows, in Section 2, we present Einstein´s field equations as an equivalent set of differential equations using a transformations due to Durgapal and Bannerji [20] . In Section 3, we make a particular choice of gravitational potential Z(x) that allows solving the field equations and we have obtained new models for charged isotropic matter. In Section 4, a physical analysis of the new solutions is performed. Finally in Section 5, we conclude.

## 2. Einstein field equations

Consider a spherically symmetric four dimensional space time whose line element is given in Schwarzschild coordinates by

$$ds^2 = -e^{2\nu(r)}dt^2 + e^{2\lambda(r)}dr^2 + r^2(d\theta^2 + \sin^2\theta d\varphi^2) \qquad (1)$$

Using the transformations, $x = Cr^2$, $Z(x) = e^{-2\lambda(r)}$ and $A^2 y^2(x) = e^{2\nu(r)}$ with arbitrary constants A and c, suggested by Durgapal and Bannerji [20], the Einstein field equations as given in (1) are

$$\frac{1-Z}{x} - 2\dot{Z} = \frac{\rho}{C} + \frac{E^2}{2c} \qquad (2)$$

$$4Z\frac{\dot{y}}{y} - \frac{1-Z}{x} = \frac{p_r}{C} - \frac{E^2}{2C} \qquad (3)$$

$$4Zx^2\ddot{y} + 2\dot{Z}x^2\dot{y} + \left(\dot{Z}x - Z + 1 - \frac{E^2 x}{C}\right)y = 0 \qquad (4)$$

$$\sigma^2 = \frac{4CZ}{x}(x\dot{E}+E)^2 \qquad (5)$$

Where $\rho$ is the energy density, $p_r$ is the radial pressure, $E$ is electric field intensity, $\sigma$ is the charge density and dot denote differentiations with respect to x. The equation (4) in the condition of pressure isotropy. We can replace the system of field equations, including the bag equation of state by the system

$$\rho = 3p + 4B \qquad (6)$$

$$\frac{p}{C} = Z\frac{\dot{y}}{y} - \frac{1}{2}\dot{Z} - \frac{B}{C} \qquad (7)$$

$$\frac{E^2}{2C} = \frac{1-Z}{x} - 3Z\frac{\dot{y}}{y} - \frac{1}{2}\dot{Z} - \frac{B}{C} \qquad (8)$$

$$\sigma = 2\sqrt{\frac{CZ}{x}}(x\dot{E}+E)^2 \qquad (9)$$

The equations (5), (6), (7), (8) and (9) governs the gravitational behavior of a charged quark star.

### 3. Generating new exact solutions

Using the procedure suggested by Komathiraj and Maharaj [9], it is possible to obtain a exact solution of the Einstein-Maxwell system. In this paper, motivated by Thirukkanesh and Ragel [12], we take the form of the gravitational potential $Z(x)$ as $Z(x) = (1-ax)^n$, where is a real constant and n is an adjustable parameter. This potential is regular at the origin and well behaved in the interior of the sphere. For the electric field we make the choice

$$E^2 = \frac{nx}{(1+ax)^2} \qquad (9)$$

This electric field is finite at the centre of the star and remains continuous in the interior. We have considered the particular cases for n=1,2, 3.

For the case n=1, the substitution of $Z(x)$ and (9) in (8), it allows to obtain the differential equation of the first order

$$\frac{\dot{y}}{y} = -\frac{x}{6C(1+ax)^2} - \frac{D}{3(1-ax)} \tag{10}$$

Where $D = \frac{B}{C} - \frac{3}{2}a$

Integrating (10), we obtain

$$y(x) = C_1(1+ax)^{\frac{-1}{24Ca^2}}(-1+ax)^{\frac{1+8DCa}{24a^2C}}\exp\left(\frac{-1}{12Ca^2(1+ax)}\right) \tag{11}$$

The equation (11) and $Z(x)$ allows generate the following analytical model:

$$e^{2\nu(r)} = C_1^2 A^2 (1+ax)^{\frac{-1}{12Ca^2}}(-1+ax)^{\frac{1+8DCa}{12a^2C}}\exp\left(\frac{-1}{6Ca^2(1+ax)}\right) \tag{12}$$

$$e^{2\lambda(r)} = \frac{1}{1-ax} \tag{13}$$

$$p = \frac{(1-ax)^2}{24a(1+ax)^2} - \frac{1+8DCa}{24a} + \frac{Ca-2B}{2} \tag{14}$$

$$\rho = \frac{(1-ax)^2}{8a(1+ax)^2} - \frac{(1+8DCa)}{8a} + \frac{3Ca}{2} + B \tag{15}$$

$$\sigma^2 = \frac{C(1-ax)(3+ax)^2}{(1+ax)^4} \tag{16}$$

With n=2, the eq. (8) becomes

$$\frac{\dot{y}}{y} = -\frac{x}{3C(1-ax)^2(1+ax)^2} - \frac{(2a^2x+F)}{3(1-ax)^2} \tag{17}$$

for convenience we have let $F = \dfrac{B}{C} - 3a$

Integrating (17), we have

$$y(x) = C_2 \dfrac{\exp\left[\dfrac{(F+2a)2Ca^2x + 4Ca^2 + 2CFa + 1}{6(-1+ax)a^2c(1+ax)}\right]}{(-1+ax)^{2/3}} \qquad (18)$$

Therefore with eq.(17) and $Z(x)$ we can generate the analytical model:

$$e^{2\nu(r)} = C_2^2 A^2 \dfrac{\exp\left[\dfrac{(F+2a)2Ca^2x + 4Ca^2 + 2CFa + 1}{3(-1+ax)a^2C(1+ax)}\right]}{(-1+ax)^{4/3}} \qquad (19)$$

$$e^{2\lambda(r)} = \dfrac{1}{(1-ax)^2} \qquad (20)$$

$$p = C(1-ax)^2 \left[\begin{array}{c} -\dfrac{2a}{3(-1+ax)} + \dfrac{(4Ca^3 + 2CFa^2)}{6a^2C(-1+ax)(1+ax)} \\ -\dfrac{((F+2a)2Ca^2x + 4Ca^2 + 2CFa + 1)}{6(-1+ax)^2 ac(1+ax)} \\ -\dfrac{((F+2a)2Ca^2x + 4Ca^2 + 2CFa + 1)}{6(-1+ax)ac(1+ax)^2} \end{array}\right] + aC(1-ax) - B \qquad (21)$$

$$\rho = 3C(1-ax)^2 \left[\begin{array}{c} -\dfrac{2a}{3(-1+ax)} + \dfrac{(4Ca^3 + 2CFa^2)}{6a^2C(-1+ax)(1+ax)} \\ -\dfrac{((F+2a)2Ca^2x + 4Ca^2 + 2CFa + 1)}{6(-1+ax)^2 ac(1+ax)} \\ -\dfrac{((F+2a)2Ca^2x + 4Ca^2 + 2CFa + 1)}{6(-1+ax)ac(1+ax)^2} \end{array}\right] + 3aC(1-ax) + B \qquad (22)$$

$$\sigma^2 = \frac{2C(1-ax)^2(3+ax)^2}{(1+ax)^4} \qquad (23)$$

With n=3, the expressions for $e^{2\nu(r)}$, $e^{2\lambda(r)}$, $p$, $\rho$ and $\sigma$ are given for

$$e^{2\nu(r)} = A^2 C_3^2 (-1+ax)^{-\frac{(3+128Ca^2)}{48Ca^2}} (1+ax)^{\frac{1}{16Ca^2}}$$
$$\exp\left[\frac{(32Ca^4 - 3a^2)x^2 - (16Ca^3 + 8CH - 3a)x - 8CHa - 6 - 48Ca^2}{24Ca^2(-1+ax)^2(1+ax)}\right] \qquad (24)$$

$$e^{2\lambda(r)} = \frac{1}{(1-ax)^3} \qquad (25)$$

$$p = C(1-ax)^3 \begin{bmatrix} -\frac{(3+128Ca^2)}{96aC(-1+ax)} + \frac{1}{32aC(1+ax)} \\ \frac{(64Ca^4 - 6a^2)x - 16Ca^3 - 8CHa^2 + 3a}{48Ca^2(-1+ax)^2(1+ax)} \\ -\frac{(32Ca^4 - 3a^2)x^2 + (3a - 16Ca^3 - 8CHa^2)x - 48Ca^2 - 8CHa - 6}{24aC(-1+ax)^3(1+ax)} \\ -\frac{(32Ca^4 - 3a^2)x^2 + (3a - 16Ca^3 - 8CHa^2)x - 48Ca^2 - 8CHa - 6}{48aC(-1+ax)^2(1+ax)^2} \end{bmatrix}$$
$$+ \frac{3aC}{2}(1-ax)^2 - B \qquad (26)$$

$$\rho = 3C(1-ax)^3 \begin{bmatrix} -\dfrac{(3+128\,Ca^2)}{96\,aC(-1+ax)} + \dfrac{1}{32\,aC(1+ax)} \\ \dfrac{(64\,Ca^4 - 6a^2)x - 16\,Ca^3 - 8CHa^2 + 3a}{48\,Ca^2(-1+ax)^2(1+ax)} \\ -\dfrac{(32\,Ca^4 - 3a^2)x^2 + (3a - 16\,Ca^3 - 8CHa^2)x - 48\,Ca^2 - 8CHa - 6}{24\,aC(-1+ax)^3(1+ax)} \\ -\dfrac{(32\,Ca^4 - 3a^2)x^2 + (3a - 16\,Ca^3 - 8CHa^2)x - 48\,Ca^2 - 8CHa - 6}{48\,aC(-1+ax)^2(1+ax)^2} \end{bmatrix}$$
$$+ \dfrac{9aC}{2}(1-ax)^2 + B \qquad (27)$$

$$\sigma^2 = \dfrac{3C(1-ax)^3(3+ax)^2}{(1+ax)^4} \qquad (28)$$

Again for convenience we have let $H = -\dfrac{9}{2}a + \dfrac{B}{C}$

### 4. Physical features of the solutions

The showed models satisfy the system of equations (6) - (9) and constitute another new family of solutions for a charged quark star with isotropic pressure. The metric functions $e^{2\nu(r)}$ and $e^{2\lambda(r)}$ can be written in terms of polinominal functions, and the variables energy density, pressure and charge density also are represented analytical. For the case n=1, The function $e^{2\nu(r)}$ is constant and behaves well inside the star and have a finite value of $e^{2\nu(r)} = C_1^2 A^2 (-1)^{\frac{1+8DCa}{12a^2C}} \exp\left(\dfrac{-1}{6Ca^2}\right)$ at the center x=0. The energy density is positive in the interior and in the center takes the value $\rho = 3aC$. The pressure p is regular and in the center x = 0 takes the value $p = aC - \frac{4B}{3}$. The charge density is continues in the interior and it vanishes in the center.

For the solution found with n=2, the functions $e^{2\nu(r)}$ and $e^{2\lambda(r)}$ acquire finite values in the center x = 0 as in the case for n=1. The energy density and the pressure take the values of $\rho = 6aC$ and $p = 2aC - \frac{4B}{3}$ at the center x=0, respectively.

With n=3, $e^{2\nu(r)}$ and $e^{2\lambda(r)}$ also have finite values at the center x=0. The energy density is $\rho = 9aC$ and the pressure $p = 3aC - \frac{4B}{3}$ at x=0. In all the cases, the charge density is continues and behaves well inside of the star.

The fact that the functions $e^{2\nu(r)}$, $e^{2\lambda(r)}$, $p$, $\rho$ and $\sigma$ have a finite value in x = 0 implies that the solutions (11) - (28) for charged quarks stars are physically acceptable and do not present singularities in the origin, as it established Jotania and Tikekar [21].

## 5. Conclusion

We have generated a new class of exact solutions for the Einstein-Maxwell system which not present singularities at the origin. We have studied three new types of analytical solutions specifying the form of the gravitational potential $Z(x)$ and of the electrical field $E$. Three obtained solutions correspond to models which have finite values for the energy density, the pressure and the charge density at the center of the star. All the solutions present a behavior similar to the solution proposed by Komathiraj and Maharaj [9] and Malaver [10] with isotropic pressure. The method to generate analytical exact solutions will depend on the form of $Z(x)$ and the electric field intensity $E$, necessary to determine physically acceptable solutions.